\documentclass[a4paper,fleqn,usenatbib]{mnras}

\usepackage[T1]{fontenc}
\usepackage{ae,aecompl}

\usepackage{graphicx}	% Including figure files
\usepackage{amsmath}	% Advanced maths commands
\usepackage{amssymb}	% Extra maths symbols
\usepackage{mathrsfs}                %Ó¢ÎÄ»¨Ìå
\usepackage{booktabs}
\usepackage{threeparttable}

\title[The Luminosity Function and Formation Rate of Long GRBs]{The Luminosity Function and Formation Rate of A Complete Sample of Long Gamma-Ray Bursts}

\author[Lan et al.]{
Guang-Xuan Lan,$^{1,2}$\thanks{E-mail: gxlan@pmo.ac.cn (GXL)}
Hou-Dun Zeng,$^{1}$\thanks{E-mail: zhd@pmo.ac.cn (ZHD)}
Jun-Jie Wei$^{1}$\thanks{E-mail: jjwei@pmo.ac.cn (JJW)}
and Xue-Feng Wu$^{1,2}$\thanks{E-mail: xfwu@pmo.ac.cn (XFW)}
\\
$^{1}$Purple Mountain Observatory, Chinese Academy of Sciences, Nanjing 210034, China\\
$^{2}$School of Astronomy and Space Science, University of Science and Technology of China, Hefei 230026, China
}

\date{Accepted XXX. Received YYY; in original form ZZZ}

\pubyear{2018}

\begin{document}
\label{firstpage}
\pagerange{\pageref{firstpage}--\pageref{lastpage}}
\maketitle

% Abstract of the paper
\begin{abstract}
We study the luminosity function and formation rate of long gamma-ray bursts (GRBs) by using a maximum likelihood method.
This is the first time this method is applied to a well-defined sample of GRBs that is complete in redshift.
The sample is composed of 99 bursts detected by the $Swift$ satellite, 81 of them with measured redshift
and luminosity for a completeness level of $82\%$. We confirm that a strong redshift evolution in luminosity
(with an evolution index of $\delta=2.22^{+0.32}_{-0.31}$) or in density ($\delta=1.92^{+0.20}_{-0.21}$) is needed in order to reproduce
the observations well. But since the predicted redshift and luminosity distributions in the two
scenarios are very similar, it is difficult to distinguish between these two kinds of evolutions only on the basis of
the current sample. Furthermore, we also consider an empirical density case in which the GRB rate density
is directly described as a broken power-law function and the luminosity function is taken to be non-evolving.
In this case, we find that the GRB formation rate rises like $(1+z)^{3.85^{+0.48}_{-0.45}}$ for $z\la2$ and is
proportional to $(1+z)^{-1.07^{+0.98}_{-1.12}}$ for $z\ga2$. The local GRB rate is $1.49^{+0.63}_{-0.64}$ Gpc$^{-3}$ yr$^{-1}$.
The GRB rate may be consistent with the cosmic star formation rate (SFR) at $z\la2$, but shows an enhancement
compared to the SFR at $z\ga2$.
\end{abstract}

% Select between one and six entries from the list of approved keywords.
% Don't make up new ones.
\begin{keywords}
gamma-ray burst: general -- stars: formation -- methods: statistical
\end{keywords}

%%%%%%%%%%%%%%%%%%%%%%%%%%%%%%%%%%%%%%%%%%%%%%%%%%

%%%%%%%%%%%%%%%%% BODY OF PAPER %%%%%%%%%%%%%%%%%%

\section{Introduction}
\label{sec:Intro}
Gamma-ray bursts (GRBs) are the most energetic explosions in the universe, which can be detected up to extremely high
redshifts. In theory, long GRBs with durations $T_{\rm 90}>2$ s (where $T_{\rm 90}$ is the time interval observed to
contain $90\%$ of the prompt emission; \citealt{1993ApJ...413L.101K}) are believed to originate from the core collapse
of massive stars (e.g., \citealt{1993AAS...182.5505W,1998ApJ...494L..45P,2006ARA&A..44..507W}), an idea given significant
support from some confirmed associations between long GRBs and supernovae (e.g., \citealt{2003Natur.423..847H,2003ApJ...591L..17S}).
This collapsar model implies that the GRB formation rate should in principle trace the cosmic star formation rate
(SFR; \citealt{1997ApJ...486L..71T,1998MNRAS.294L..13W,2000ApJ...536....1L,2001ApJ...548..522P,2004RvMP...76.1143P,2004IJMPA..19.2385Z,2007ChJAA...7....1Z}).
However, the $Swift$ observations seem to indicate that the GRB rate does not closely follow the SFR but is actually enhanced
by some unknown mechanisms at high-$z$ \citep{2006ApJ...647..773D,2007JCAP...07..003G,2007ApJ...661..394L,
2007ApJ...656L..49S,2008ApJ...673L.119K,2009ApJ...705L.104K,2008MNRAS.388.1487L,2008ApJ...683L...5Y,
2009MNRAS.396..299S,2012ApJ...749...68S,2010MNRAS.407.1972C,2010MNRAS.406..558Q,2010MNRAS.406.1944W,
2011MNRAS.416.2174C,2011MNRAS.417.3025V,2012A&A...539A.113E,2012ApJ...745..168L,2012ApJ...744...95R,
2013ApJ...772L...8T,2013A&A...556A..90W,2014MNRAS.439.3329W,2015MNRAS.454.1785T,2016ApJ...820...66D,
2017IJMPD..2630002W,2018MNRAS.473.3385P}.\footnote{Using the $C^{-}$ statistical method proposed by \cite{1971MNRAS.155...95L},
\cite{2015MNRAS.447.1911P} and \cite{2015ApJS..218...13Y} found a relative excess of the GRB formation rate
with respect to the SFR at low redshifts. But then \cite{2016A&A...587A..40P} showed that if the $C^{-}$ method
is applied to incomplete GRB samples it can misleadingly lead to an excess of the GRB rate at $z\leq1$.
In addition, some works performed spectro-photometric studies on the properties (stellar
mass, SFR, and metallicity) of long GRB host galaxies of different complete GRB samples and
compared them to the ones of typical star-forming galaxies selected by galaxy surveys (e.g.,
\citealt{2015A&A...581A.102V,2016A&A...590A.129J,2016ApJ...817....8P,2019A&A...623A..26P}).
All their results clearly suggested that at $z\leq1$ only a small fraction of the star formation produces GRBs.}
Several evolution models have been proposed to explain the observed enhancement, such as the GRB rate
density evolution \citep{2008ApJ...673L.119K,2009ApJ...705L.104K}, cosmic metallicity evolution
\citep{2006ApJ...638L..63L,2008MNRAS.388.1487L}, and an evolution in the GRB luminosity function
\citep{2011MNRAS.417.3025V,2012ApJ...749...68S,2013ApJ...772L...8T,2015MNRAS.454.1785T,2018MNRAS.473.3385P}.

However, it should be emphasized that our knowledge about the properties of long GRBs and their evolution
with cosmic time is still hindered by the fact that most of the observed $Swift$ GRBs are without redshift.
Indeed, only $\sim1/3$ of all $Swift$ GRBs have redshift determinations. Most of previous researches adopted
incomplete redshift samples to derive the GRB luminosity function and redshift distribution. Given the low completeness level in redshift
measurement, the possible observational biases may have remarkable effect on sharping the GRB redshift distributions
\citep{2007A&A...470..515F,2012ApJ...749...68S}. Therefore, it is necessary to consider using an unbiased
complete sample of long GRBs that is capable of adequately representing this class of object to study
their distributions through cosmic time. \cite{2012ApJ...749...68S} defined a complete flux-limited sample of
$Swift$ long GRBs which, despite containing a relatively small sample size, has a completeness level
in redshift determination of 90\%. The high level of redshift completeness enabled them for the first time
to constrain the GRB luminosity function and its evolution in an unbiased way. They found that either a luminosity evolution
with $\delta=2.1\pm0.6$ or a density evolution with $\delta=1.7\pm0.5$ can well reproduce the observations.
However, they can not discriminate between these two scenarios. Recently, \cite{2016A&A...587A..40P} revised the complete
sample of \cite{2012ApJ...749...68S} and then extended it with new bursts that have favorable observing conditions
for redshift determination and that are bright in the 15--150 keV $Swift$/BAT band. The updated sample is composed of
99 bursts, 81 of them with known redshift and luminosity for a completeness level of 82\%. \cite{2016A&A...587A..40P} adopted
the Lynden-Bell $C^{-}$ method to derive the luminosity function and formation rate of GRBs from this updated complete $Swift$
sample. A strong evolution in luminosity $L(z)\propto (1+z)^{2.5}$ was found.

There are several algorithms to derive the luminosity function for a specific kind of astronomical sources.
A classical approach to determine the luminosity function is based on the $1/V_{\rm MAX}$ method of
\cite{1968ApJ...151..393S} applied to redshift bins. However, it is known that this method would
introduce bias if there is strong evolution within the bins. Moreover, given the relatively small number
of GRBs and their wide redshift and luminosity range spanned, binning would lead to a loss of information.
Other non-parametric methods (e.g., the $C^{-}$ method; \citealt{1971MNRAS.155...95L}) usually require certain
uniformity of data coverage to be applicable. In practice, the GRB data are truncated by the flux sensitivity limit
of the detector. It is very difficult to parametrize the sensitivity of the detector and to construct a uniformly
distributed GRB sample. The maximum-likelihood algorithms \citep{1983ApJ...269...35M} are therefore
preferable for the GRB problems, because these methods are more flexible in modeling the systematical uncertainties
and are less limited by the conditions of a given sample. \cite{2010MNRAS.406.1944W} adopted a maximum likelihood
estimator to obtain the GRB luminosity function and formation rate. They analyzed a sample of $\sim100$ $Swift$ GRBs
with known redshifts, which is possibly suffering from incompleteness.

In this work, we make use of the high completeness of the updated sample presented in \cite{2016A&A...587A..40P}
to constrain the GRB luminosity function and redshift distribution. Compared with previous works using incomplete samples,
which relied on the assumption that bursts lacking redshift information strictly follow the redshift distribution
of bursts with measured redshifts, our present work has the advantage of being independent of this assumption.
Additionally, to include systematics and unknowns in the statistical inference,
we apply the maximum likelihood method, for the first time, to analyze the complete GRB sample. The rest of the paper
is organized as follows. In Section~\ref{sec:sample}, we describe the complete sample at our disposal.
In Section~\ref{sec:method}, we illustrate the maximum likelihood method used for our analysis.
Our models and analysis results are presented in Section~\ref{sec:results}. Lastly, we draw a brief summary in
Section~\ref{sec:summary}. Throughout this paper we adopt a standard $\Lambda$CDM cosmological model
with $\Omega_{m}=0.3$, $\Omega_{\Lambda}=0.7$, and $H_{0}=70$ km s$^{-1}$ Mpc$^{-1}$.

\section{The Sample}
\label{sec:sample}
Since the launch of the $Swift$ satellite \citep{2004ApJ...611.1005G}, the number of measured GRB redshifts has increased rapidly.
But we still have to face the problem that the bursts with redshift measurements account for only $\sim1/3$ of all GRBs.
The low completeness level in redshift determination will undoubtedly lead to biases in the statistical analysis of GRBs.
\cite{2006A&A...447..897J} thus proposed some criteria to select long GRBs which have favorable observing conditions for
redshift measurement. \cite{2012ApJ...749...68S} built a complete sample of $Swift$ long GRBs (called BAT6), which is composed of 58 GRBs
matching the criteria of \cite{2006A&A...447..897J} and having 1-s peak photon flux $P\ge 2.6$ ph cm$^{-2}$ s$^{-1}$ (integrated in
the 15--150 keV BAT energy band). 52 of them have measured redshift so that the completeness level is $\sim90$\%.
\cite{2016A&A...587A..40P} revised the BAT6 sample and extended it with additional bursts that satisfy its selection
criteria. The BAT6 extended (BAT6ext) sample contains 99 GRBs up to 2014 July, of which 81 bursts have measured $z$
and $L$. Its completeness in redshift is $\sim82$\%. In the following, we will use the BAT6ext sample
to investigative the GRB luminosity function and redshift distribution.

\section{Maximum Likelihood Analysis}
\label{sec:method}
In order to constrain the model parameters, a maximum likelihood method first introduced by \cite{1983ApJ...269...35M} is adopted.
%In this method, the luminosity--redshift plane is parsed into extremely small intervals of size $dLdz$.
%Assuming that the number of observed bursts ($x$) within each $dLdz$ element and a certain observational time
%obey the Poisson distribution, we can define a likelihood function:
%\begin{equation}
%f(x)=\frac{e^{-\mu}\mu^x}{x!}\;,
%\end{equation}
%where $\mu$ is the expected number of bursts. If $x=1$ then the function is ${\mu}{e^{-\mu}}$.
The likelihood function $\mathcal{L}$ is defined by the expression
\citep{1998ApJ...496..752C,2006ApJ...643...81N,2009ApJ...699..603A,2012ApJ...751..108A,2010ApJ...720..435A,2014MNRAS.441.1760Z,2016MNRAS.462.3094Z}
\begin{equation}
\mathcal{L}=\exp(-N_{\rm exp})\prod ^{N_{\rm obs}}_{i=1} \Phi(L_i,z_i,t_i)\;,
\end{equation}
where $N_{\rm exp}$ is the expected number of GRB detections, $N_{\rm obs}$ is the number of the observed sample,
and $\Phi(L,z,t)$ is the observed rate of GRBs per unit time at redshift $\in(z, z+dz)$ with luminosity
$\in(L, L+dL)$, which can be expressed as
\begin{equation}
\Phi\left(L,z,t\right)=\frac{d^3 N}{dtdzdL}=\frac{\Delta \Omega}{4\pi}\frac{\psi(z)}{1+z}\frac{dV(z)}{dz}\phi(L)\;,
\end{equation}
where $\Delta \Omega=1.33$ sr is the solid angle covered on the sky by $Swift$, $\psi(z)$ is the comoving formation rate
of GRBs in units of $\rm Gpc^{-3}$ $\rm yr^{-1}$, $(1+z)^{-1}$ is the cosmic expansion factor, and $\phi(L)$ is the normalized GRB luminosity function.
And $dV(z)/dz=4\pi c D_L^2(z)/[H(z)(1+z)^2]$ is the comoving volume element, where $D_L(z)$ is the luminosity distance
and $H(z)=H_0[\Omega_{m}(1+z)^3+\Omega_{\Lambda}]^{1/2}$ is the Hubble parameter.
Transforming the likelihood function to the standard expression $\chi^2=-2\ln L$, we obtain the $\chi^2$ distribution
for the complete $Swift$ sample:
\begin{equation}
\chi^2_{\rm Swift}=-2\sum^{N_{\rm obs}}_{i=1} \ln\left[\Phi(L_i,z_i,t_i)\right]+2N_{\rm exp}\;.
\end{equation}
Considering the flux threshold used to defined the BAT6ext (i.e., $P_{\rm lim}=2.6$ ph cm$^{-2}$ s$^{-1}$ in the 15--150 keV energy band),
the expected number of GRBs can be estimated by
\begin{equation}
N_{\rm exp} = \frac{\Delta \Omega T}{4\pi} \int ^{z_{\rm max}} _{0}  \frac{\psi(z)}{1+z} \frac{dV(z)}{dz}dz\int ^{L_{\rm max}}_{{\rm max}[L_{\rm min},L_{\rm lim}(z)]}\phi(L)dL\;,
\label{eq:Nexp}
\end{equation}
where $T\sim9.3$ yr is the observational period of $Swift$ that covers the BAT6ext sample.
Since $z < 10$ for the current GRB sample, we adopt a maximum GRB redshift $z_{\rm max}=10$.
The luminosity function is assumed to extend between minimum and maximum luminosities $L_{\rm min}=10^{49}$
erg $\rm s^{-1}$ and $L_{\rm max}=10^{55}$ erg $\rm s^{-1}$ \citep{2015MNRAS.447.1911P}.
The luminosity threshold appearing in Equation~(\ref{eq:Nexp}) can be calculated by
\begin{equation}
L_{\rm lim}(z)=4\pi D_L^2(z) P_{\rm lim} \frac{\int^{10^4/(1+z)\;{\rm keV}}_{1/(1+z)\;{\rm keV}} EN(E)dE}{\int^{150\;{\rm keV}}_{15\;{\rm keV}} N(E)dE}\;,
\end{equation}
where $N(E)$ is the observed photon spectrum. To describe the typical GRB spectrum, we use a Band function
with low- and high-energy spectral indices $-1$ and $-2.3$, respectively \citep{1993ApJ...413..281B,2000ApJS..126...19P,2006ApJS..166..298K}.
The spectral peak energy $E_{p}$ is obtained through the $E_{p}$--$L$ correlation \citep{2004ApJ...609..935Y,2012MNRAS.421.1256N}:
$\log \left[E_{p}(1+z)\right]=-25.33+0.53\log L$, where $L$ represents the isotropic peak luminosity.

Similar to what \cite{2012ApJ...749...68S} did in their treatment, we optimize the model free parameters by jointly fitting
the observed redshift and luminosity distributions of bursts in the BAT6ext sample and the observed differential
peak-flux number counts in the 50--300 keV band of $Fermi$/GBM \citep{2014ApJS..211...12G,2014ApJS..211...13V,2016ApJS..223...28N}.\footnote{
The $Fermi$/GBM GRB data are available at https://fermi.gsfc.nasa.gov/ssc/data/access/gbm/.} The $Fermi$/GBM sample
contains $\sim10.8$ yr of observation (up to 2019 May) with an average exposure factor of $\sim0.5$. To avoid the
complication that would arise from the use of a detailed treatment of the $Fermi$/GBM threshold, we select 1375 $Fermi$/GBM
long GRBs with peak flux $P\ge 1.0$ ph cm$^{-2}$ s$^{-1}$ in the 50--300 keV energy band.
The expected number of events in each peak-flux bin $P\in(P_1,P_2)$ should be
\begin{equation}
\begin{aligned}
N^{\rm exp}(P_1<P<P_2) = 0.5\times 10.8&\times  \int ^{z_{\rm max}} _{0}  \frac{\psi(z)}{(1+z)} \frac{dV(z)}{dz}\,dz\\
&\times\int ^{L(P_2,z)}_{L(P_1,z)} \phi(L) \,dL\;.
\end{aligned}
\end{equation}
The $\chi^2$ value for the $Fermi$ sample is then given by
\begin{equation}
\chi^2_{\rm Fermi}=\sum_{i}^{n}\frac{\left(N^{\rm obs}_{i}-N^{\rm exp}_{i}\right)^2}{\sigma^{2}_{i}}\;,
\end{equation}
where $n$ is the number of $P$ bins, and $N^{\rm obs}$ and $\tilde{N}^{\rm exp}$ are the observed and expected numbers of GRBs in bin $i$, respectively.
For the observed number $N^{\rm obs}_{i}$ in bin $i$, the statistical error of $N^{\rm obs}_{i}$ is usually
considered to be the Poisson error, i.e., $\sigma_{i}=\sqrt{N^{\rm obs}_{i}}$, which denotes the 68\% Poisson
confidence intervals for the binned events. Here the differential peak-flux distribution is treated as
a sum of independent measurements in the different 20 $P$ bins with the same width $\Delta \log P$ in $\log (P)$ space
(i.e., $\Delta \log P=[\log (P_{\rm max})-\log (P_{\rm min})]/20$).
Note that taking different values for $\Delta \log P$ has little impact on the best-fitting results.
It should be emphasized that the measure of the local GRB rate density is only determined by
the observed total number of $Fermi$ GRBs detected above $P= 1.0$ ph cm$^{-2}$ s$^{-1}$.

Therefore, we can define a new $\chi^2_{\rm total}$ function that combining the maximum likelihood analysis and
the constraints from the number counts of $Fermi$:
\begin{equation}
\chi^2_{\rm total}=\chi^2_{\rm Swift}+ \chi^2_{\rm Fermi}\;.
\label{eq:chi2_total}
\end{equation}
The complete $Swift$ sample can provide a powerful test for the existence and the level of redshift evolution of long GRBs,
while the fit to the $Fermi$ number counts enables us to infer the local GRB rate density and to better constrain the luminosity function
free parameters \citep{2012ApJ...749...68S}.
For each model, we optimize the free parameters using the Markov Chain Monte Carlo (MCMC) technology,\footnote{We adopt
the MCMC code from CosRayMC \citep{2012PhRvD..85d3507L}, which itself was adopted from the COSMOMC package \citep{2002PhRvD..66j3511L}.}
which is widely employed to give multidimensional parameter constraints from the observational data.
In practice, this means we will find the parameter values that minimize $\chi^2_{\rm total}$, which
yields the best-fitting parameters and their corresponding $1\sigma$ uncertainties.
It is worth stressing that the derived best-fitting parameters also give a good fit of the $Swift$
peak-flux number counts in the 15--150 keV band.

\section{Models and Analysis Results}
\label{sec:results}

Here we explore the expression of the GRB luminosity function in a broken power law, which is widely adopted in the literature:
\begin{equation}
 \phi (L)  = \frac{A}{\ln(10)L}\left\{\begin{array}{l}
{\left(\frac{{{L}}}{{{L_c}}}\right)^a};\,\,{L} \le {L_c}\\
{\left(\frac{{{L}}}{{{L_c}}}\right)^b};\,\,{L} > {L_c}\;,
\end{array} \right.
\end{equation}
where $A$ is a normalization constant, $L_c$ is the break luminosity, and $a$ and $b$ are the faint- and bright-end
power-law indices, respectively.

\begin{figure}
  \centering
  \includegraphics[width=\columnwidth]{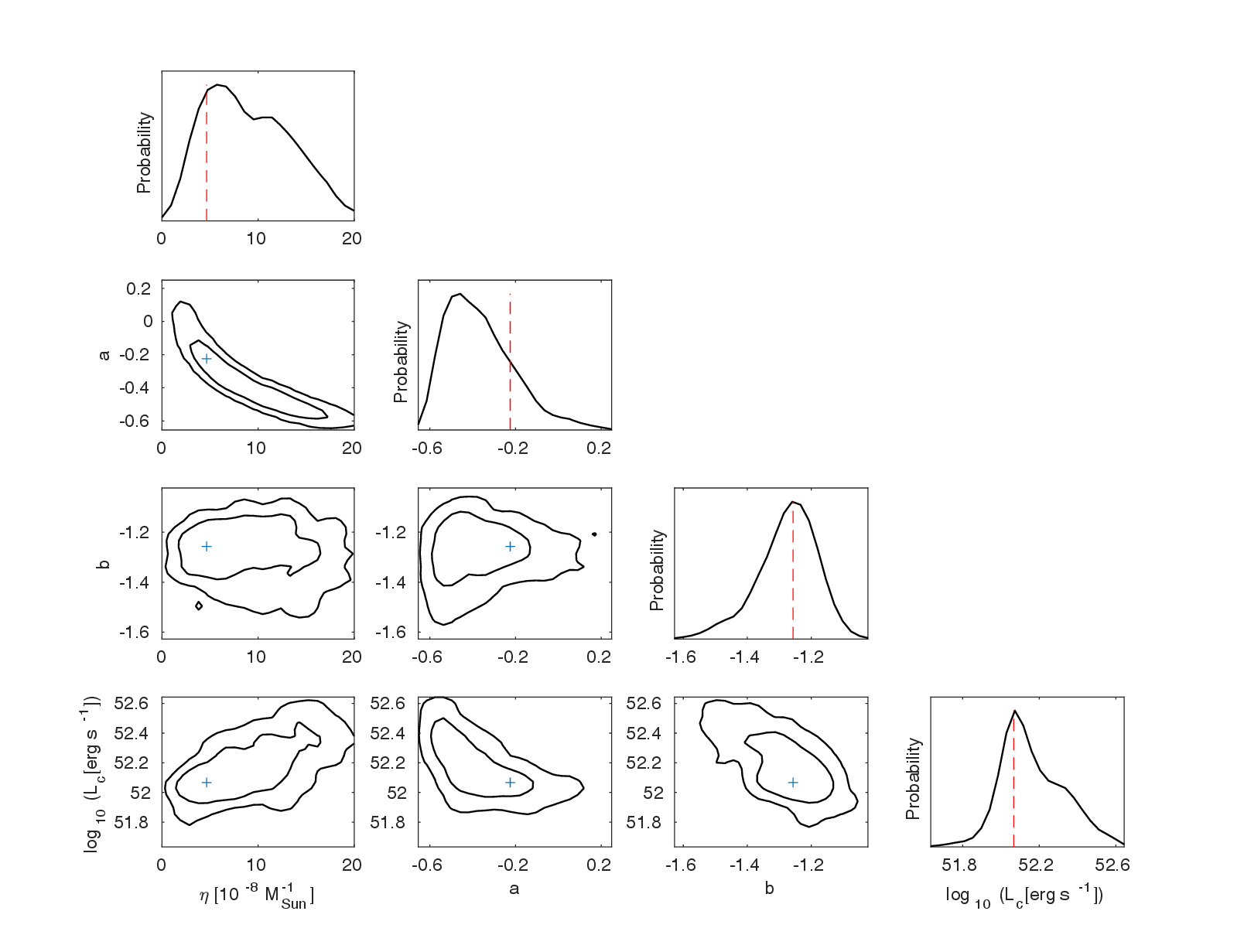}\\
  \caption{1-D probability distributions and 2-D regions with the 1--2$\sigma$ contours corresponding to the parameters
  $\eta$, $a$, $b$, and $L_{c,0}$ in the no-evolution model. The vertical dashed lines represent the best-fitting values.}
  \label{fig1}
\end{figure}

\begin{figure}
  \centering
  \scriptsize
  \includegraphics[width=\columnwidth]{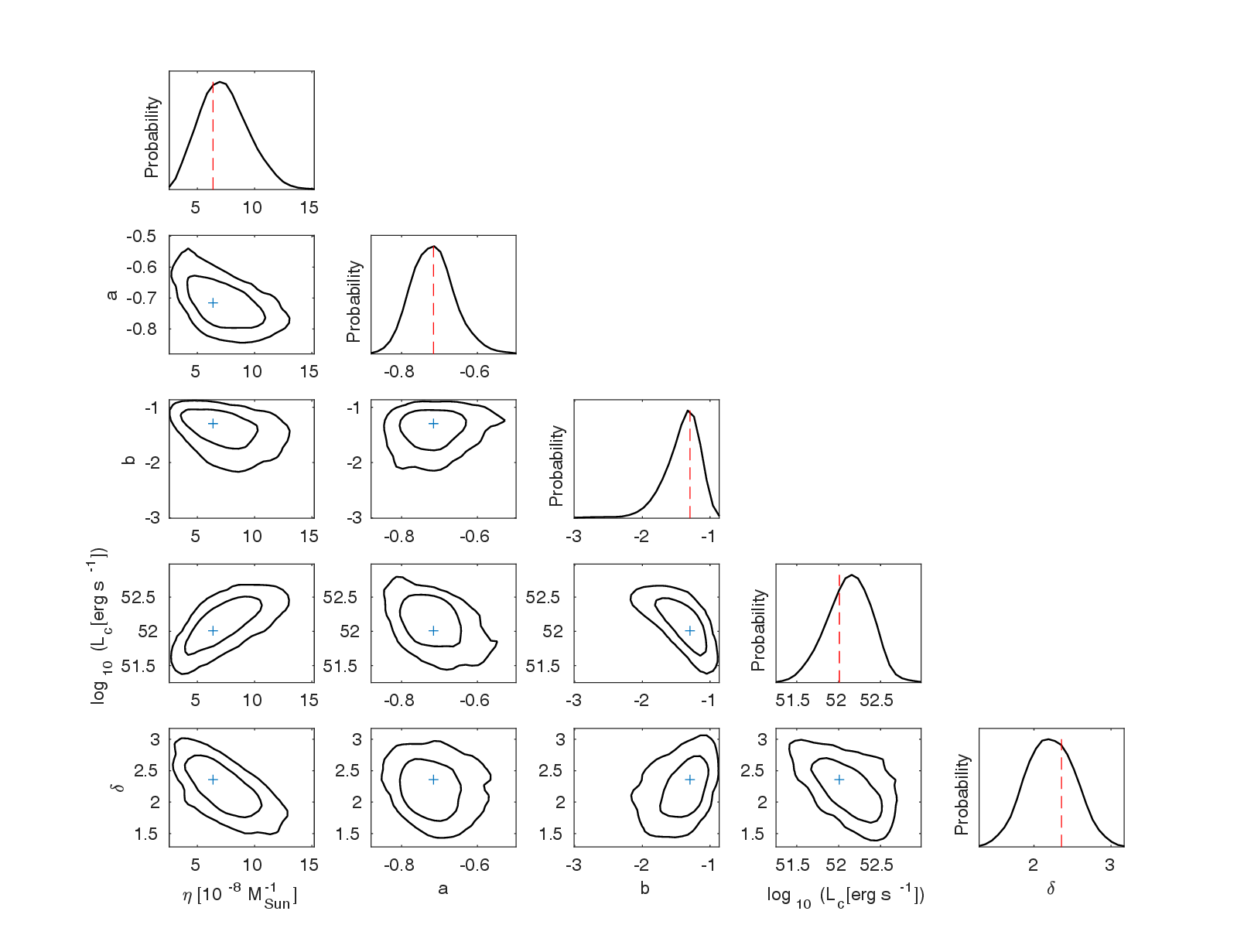}\\
  \caption{Same as Figure~\ref{fig1}, but for the parameters $\eta$, $a$, $b$, $L_{c,0}$, and $\delta$ in the luminosity evolution model.}
  \label{fig2}
\end{figure}

\begin{figure}
  \centering
  \scriptsize
  \includegraphics[width=\columnwidth]{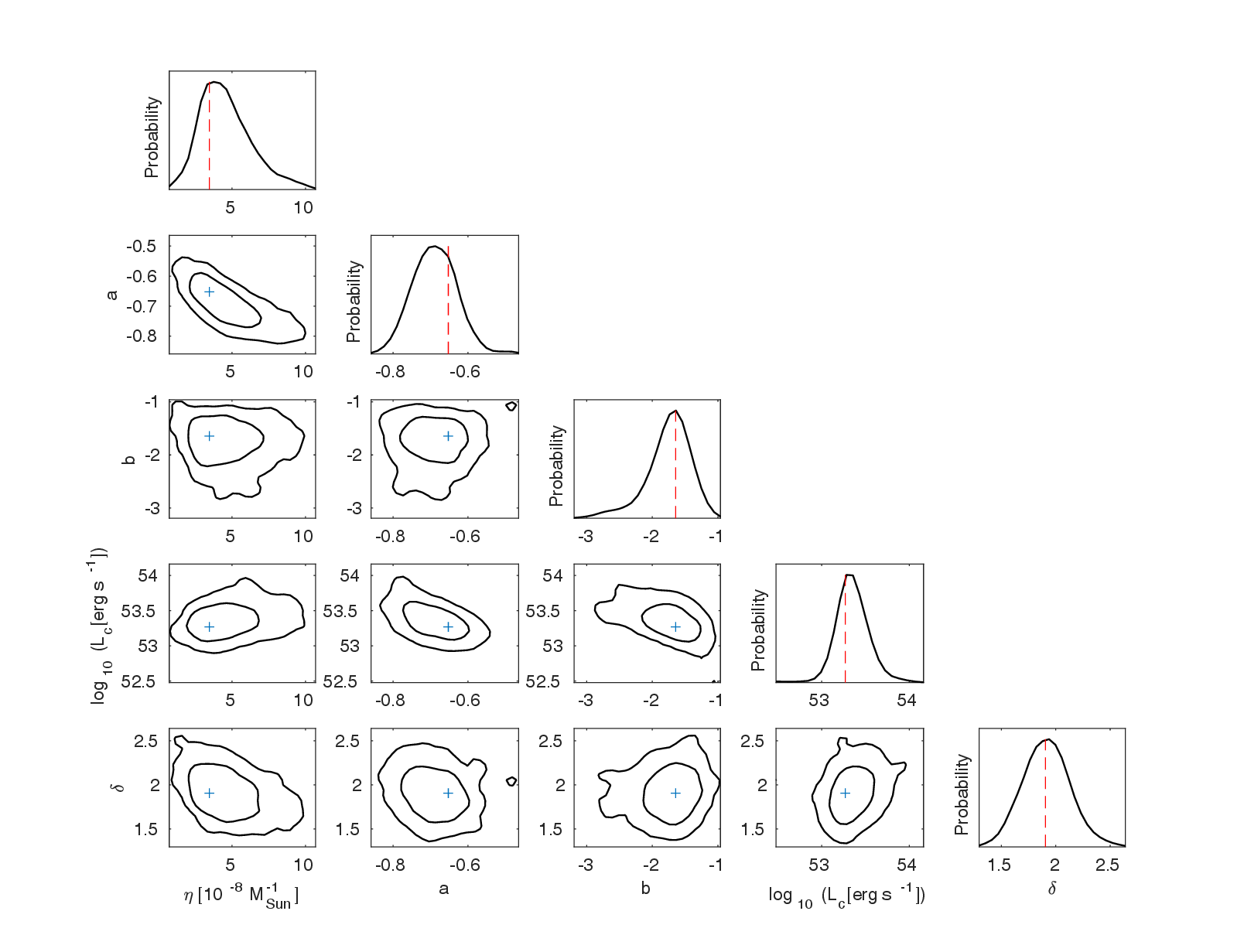}\\
  \caption{Same as Figure~\ref{fig1}, but for the parameters $\eta$, $a$, $b$, $L_{c,0}$, and $\delta$ in the density evolution model.}
  \label{fig3}
\end{figure}

\begin{figure}
  \centering
  \includegraphics[width=\columnwidth]{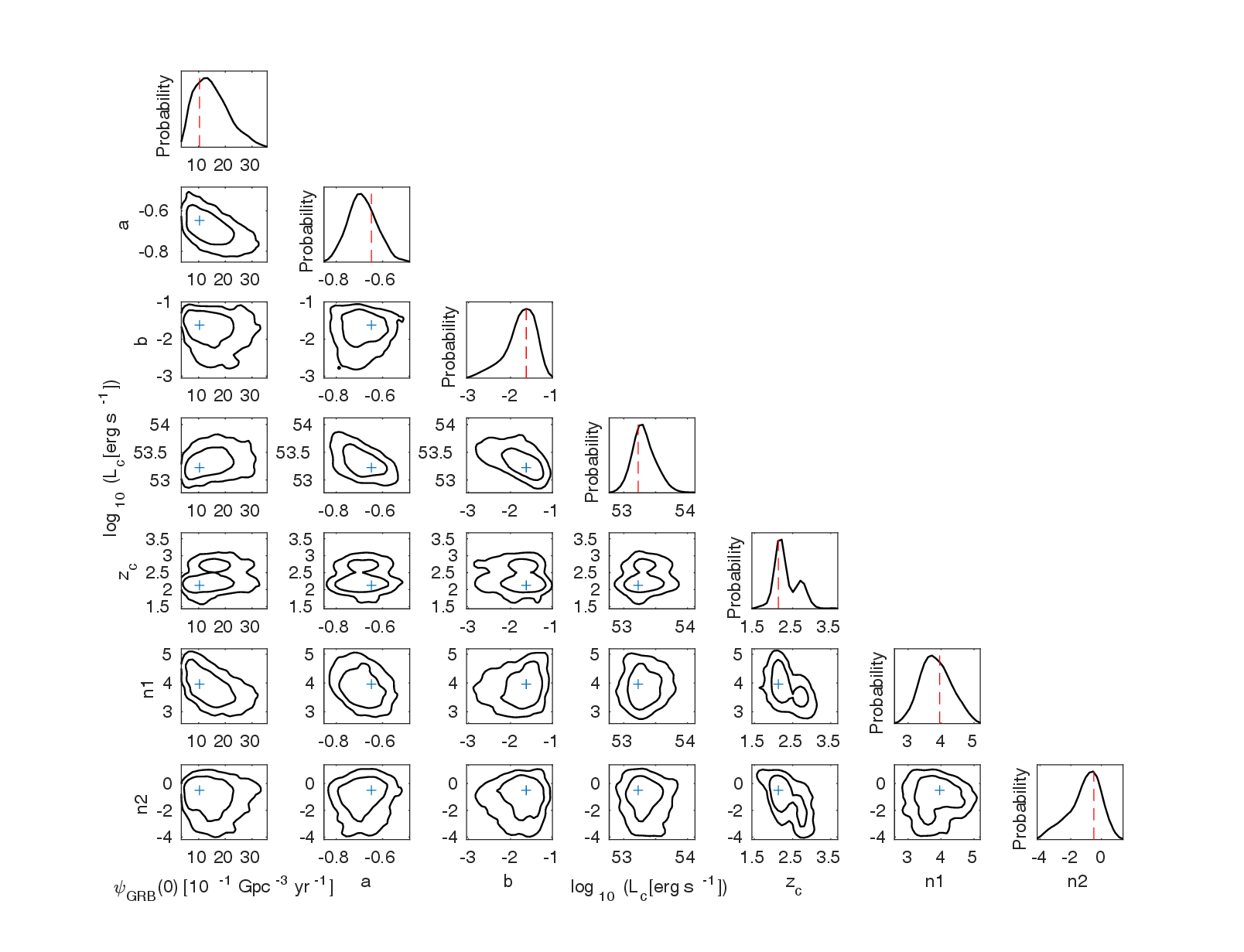}\\
  \caption{Same as Figure~\ref{fig1}, but for the parameters $\psi_{\rm GRB}(0)$, $a$, $b$, $L_{c,0}$, $z_c$, $n_1$, and $n_2$
  in the empirical density model.}
  \label{fig4}
\end{figure}

Using the maximum likelihood analysis method described in Section~\ref{sec:method}, we can optimize the values of each model's
free parameters, including the GRB luminosity function, the GRB formation efficiency $\eta$, and the evolution parameter. Table~\ref{tab1} lists
the best-fitting parameters together with their $1\sigma$ confidence level for different models. In the last two columns, we report
the total $\chi_{\rm total}^{2}$ value and the Akaike information criterion (AIC) score, respectively. For each fitted model,
the AIC is given by ${\rm AIC}\equiv\chi_{\rm total}^2+2k$, where $k$ is the number of free parameters \citep{1974ITAC...19..716A,2007MNRAS.377L..74L}.
If there are two or more models for the data, $\mathcal{M}_{\rm 1}, \mathcal{M}_{\rm 2},..., \mathcal{M}_{\rm N}$, and they have been separately
fitted, the one with the least AIC score is the one most favoured by this criterion. A more quantitative ranking of models can be
calculated as follows. With ${\rm AIC}_\alpha$ characterizing model $\mathcal{M}_\alpha$, the un-normalized confidence that this
model is true is the Akaike weight $\exp(-{\rm AIC_\alpha}/2)$. The relative probability that $\mathcal{M}_\alpha$ is statistically
preferred is
\begin{equation}
P(\mathcal{M}_\alpha)=\frac{\exp(-{\rm AIC_\alpha}/2)}{\exp(-{{\rm AIC_1}}/2)+\cdot\cdot\cdot+\exp(-{\rm AIC_N}/2)}\;.
\end{equation}
One-dimensional (1-D) probability distributions and two-dimensional (2-D) regions with the 1--2$\sigma$ contours corresponding to
the parameters in different models are shown in Figures~\ref{fig1}-\ref{fig4}. One can see from these plots that all of the model
parameters are well constrained.

\newcommand{\tabincell}[2]{\begin{tabular}{@{}#1@{}}#2\end{tabular}}
\begin{table*}
%\scriptsize
\centering
\caption{Best-fitting parameters in different models.}
\label{tab1}
\begin{tabular}{|l|c|c|c|c|c|c|c|}
\hline
  Model &  \tabincell{c}{Evolution parameter} & $\eta$ & $a$ & $b$ & $\log{L_c}$ & $\chi_{\rm total}^2$ & AIC \\
        &                      & ($10^{-8}$ ${\rm M}_{\odot}^{-1}$) &  &  & (erg $\rm s^{-1}$) &  &  \\

	  \hline
  No evolution&  \tabincell{c}{$\cdot\cdot\cdot$} & $9.00^{+2.22}_{-2.67}$ &$-0.36^{+0.16}_{-0.16}$ & $-1.28^{+0.09}_{-0.09}$ &$52.17^{+0.18}_{-0.14}$ & 20169.7 & 20177.7 \\
  Luminosity evolution &  \tabincell{c}{$\delta=2.22^{+0.32}_{-0.31}$} & $7.35^{+2.11}_{-2.06}$ &$-0.72^{+0.05}_{-0.05}$& $-1.42^{+0.24}_{-0.24}$ & $52.12^{+0.26}_{-0.26}$& 20086.1& 20096.1\\
  Density evolution &  \tabincell{c}{$\delta=1.92^{+0.20}_{-0.21}$} & $4.60^{+1.67}_{-1.63}$ &$-0.69^{+0.05}_{-0.06}$ & $-1.75^{+0.31}_{-0.29}$ & $53.35^{+0.16}_{-0.15}$& 20083.2 & 20093.2 \\
  Empirical density & \tabincell{c}{$\psi_{\rm GRB}(0)=1.49^{+0.63}_{-0.64}$\\$n_1=3.85^{+0.48}_{-0.45}$\\$n_2=-1.07^{+0.98}_{-1.12}$\\$z_c=2.33^{+0.39}_{-0.24}$} & $\cdot\cdot\cdot$ & $-0.69^{+0.06}_{-0.06}$ & $-1.76^{+0.34}_{-0.33}$ & $53.32^{+0.18}_{-0.17}$ & 20079.0 & 20093.0 \\
  \hline
\end{tabular}
\begin{description}
  \item[Note.] {The GRB formation rate at $z=0$, $\psi_{\rm GRB}(0)$, is given in units of Gpc$^{-3}$ yr$^{-1}$.
                The parameter values were calculated as the median of all the best-fitting parameters to the Monte Carlo sample, while the uncertainties correspond to the 68\% containment regions around the median values.}

\end{description}
\end{table*}

\subsection{No evolution model}
In the first simple (no-evolution) model, we assume that the GRB formation rate purely follows the cosmic SFR,
$\psi_\star(z)$, i.e., $\psi(z)=\eta \psi_\star(z)$, and that their luminosity function does not evolve with redshift, i.e.,
$L_{c,z}=L_{c,0}={\rm const}$. The factor $\eta$ denotes the GRB formation efficiency in units of ${\rm M}_{\odot}^{-1}$.
The observed SFR $\psi_\star(z)$ (in units of ${\rm M}_{\odot}$ $\rm yr^{-1}$ $\rm Mpc^{-3}$) is commonly
parameterized with the form \citep{2006ApJ...651..142H,2008MNRAS.388.1487L}:
\begin{equation}
\psi_{\star}(z)=\frac{0.0157+0.118z}{1+(z/3.23)^{4.66}}\;.
\end{equation}
The 1-D probability distributions and 2-D regions with the 1--2$\sigma$ contours corresponding to
four parameters in the no-evolution model are presented in Figure~\ref{fig1}.

\begin{figure*}
\includegraphics[angle=0,scale=0.5]{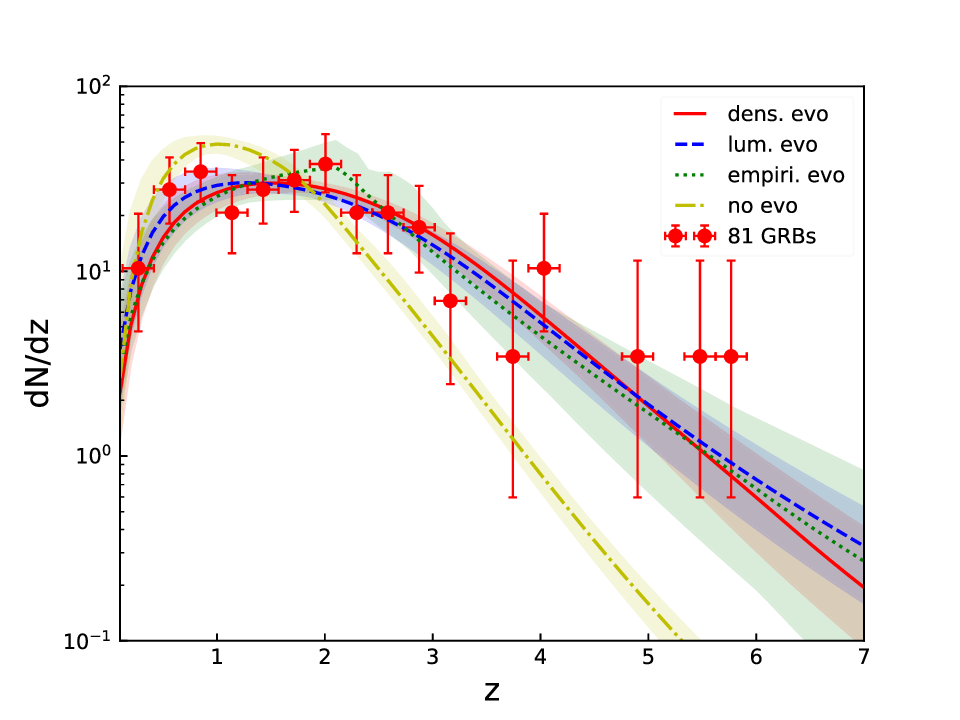} %\hskip 0.1in
\includegraphics[angle=0,scale=0.5]{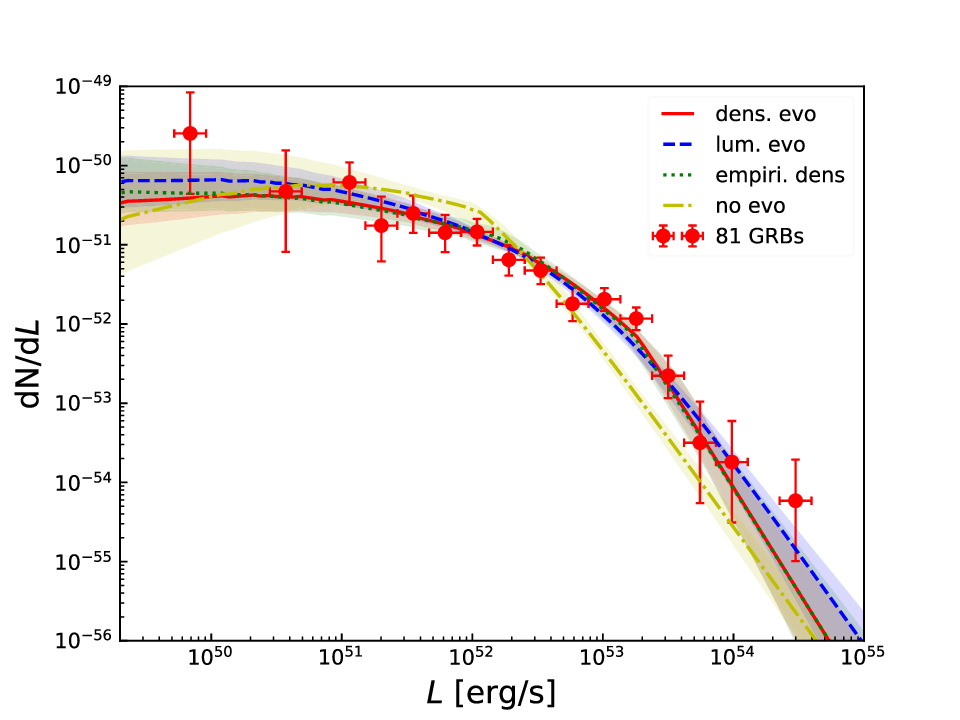}
\vskip-0.1in
    \caption{Redshift (left panel) and luminosity (right panel) distributions of GRBs with $P\ge 2.6$ ph cm$^{-2}$ s$^{-1}$ in the 15--150 keV energy band.
    Data points are the observed redshift and luminosity distributions of 81 GRBs in the complete $Swift$ sample,
    and the error bars represent the Poisson errors of the number of detection in each redshift or luminosity bin.
    Curves show the expected distributions for different best-fitting models: no-evolution model (yellow dot-dashed lines),
    luminosity evolution model (blue dashed lines), density evolution model (red solid lines), and empirical
    density model (green dotted lines). Shaded regions show the $1\sigma$ confidence regions of the corresponding models.}\label{fig5}
\end{figure*}

Figure~\ref{fig5} shows the $z$ and $L$ distributions of 81 GRBs in the BAT6ext complete sample.
The expectation from the no-evolution case (yellow dot-dashed lines) do not provide a good representation
of the observed $z$ and $L$ distributions of the BAT6ext sample. Particularly, the rate of GRBs
at high redshift is clearly under-predicted and the reproduce of the $L$ distribution is not
as good as those of the luminosity evolution model or density evolution model, more fully described below.
According to the AIC model selection criterion, we can safely discard this model as having
a probability of only $\sim10^{-19}$ of being correct compared to the other three models.

\subsection{Luminosity evolution model}
This model assumes that an evolution in the GRB luminosity function can enhance the number of GRB detections at high-\emph{z}.
In this case, while the GRB formation rate is still proportional to the cosmic SFR, the break luminosity
in the GRB luminosity function increases with redshift as $L_{c,z}=L_{c,0}(1+z)^{\delta}$.
In Figure~\ref{fig2}, we also display the 1-D probability distributions and 1--2$\sigma$ constraint contours for
five parameters in this model.
We find that a strong luminosity evolution with $\delta=2.22^{+0.32}_{-0.31}$ is required to reproduce both the observed
$z$ and $L$ distributions of 81 bursts in the BAT6ext sample (blue dashed lines in Figure~\ref{fig5}).
Based on the same sample, \cite{2016A&A...587A..40P} found a strong evolution in luminosity ($\delta=2.5$)
through the Lynden-Bell $C^{-}$ method, in good agreement with our results from the maximum likelihood analysis method.
Using the AIC model selection criterion, we find that this model is somewhat disfavored statistically
compared to the other three models, with a relative probability of $\sim10\%$.

\subsection{Density evolution model}
This model assumes that an evolution of the GRB formation rate can also provide an enhancement of the high-\emph{z}
GRB detection. In this scenario, while the break luminosity in the GRB luminosity function is still a constant,
the GRB rate follows the cosmic SFR in conjunction with an additional evolution characterized by $(1+z)^{\delta}$,
i.e., $\psi(z)=\eta \psi_\star(z) (1+z)^\delta$. The best-fitting parameters and their constraint contours
are shown in Figure~\ref{fig3}. We find that a strong density evolution with $\delta=1.92^{+0.20}_{-0.21}$
reproduces the observed $z$ and $L$ distributions (red solid lines in Figure~\ref{fig5}) quite well.
On the basis of the AIC model selection criterion, we find that among four different
models, this one is statistically preferred with a relative probability $\sim 43\%$.

The large value of $\delta$ means an obvious shift of the peak of the GRB formation rate toward a
higher redshift with respect to stars. In this following section, we will further investigate this issue
by directly adopting an empirical function as the GRB rate density, without making any assumption on the relation between
the GRB rate and the SFR. Later on we will compare this empirical function with models for the GRB rate
that follow the SFR.

\subsection{Empirical density model}
\begin{figure}
  \centering
  \includegraphics[width=\columnwidth]{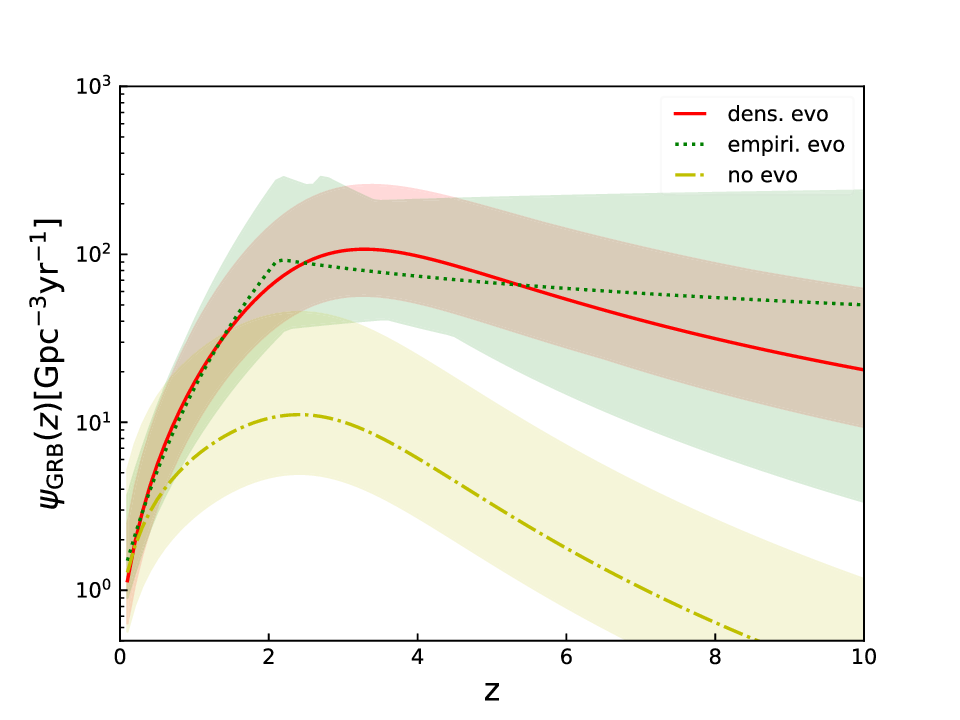}\\
  \caption{Intrinsic redshift distribution of GRBs for different models. Curves show the results for
  the no-evolution model (yellow dot-dashed line), for the density evolution model (red solid line),
  and for the empirical density model (green dotted line), respectively. The shaded regions are
  the $1\sigma$ uncertainties of the corresponding models.}
  \label{fig6}
\end{figure}

We approximate the expression of the GRB formation rate with an empirical broken power-law function \citep{2010MNRAS.406.1944W}:
\begin{equation}
  \psi(z) = {\psi_{\rm GRB}}(0)\left\{ \begin{array}{l}
{(1 + z)^{{n_1}}};\;\;\;\;\;\;\;\;\;\;\;\;\;\;\;\;\;\;\;\;\;\; z\le {z_c}\\
{(1 + {z_c})^{{n_1} - {n_2}}}{(1 + z)^{{n_2}}};\;z > {z_c}\;,
\end{array} \right.
\end{equation}
where $\psi_{\rm GRB}(0)$ is the local GRB formation rate and $z_c$ is the break redshift.
Same as the density evolution model, there is no evolution of the GRB luminosity function in this case.
Figure~\ref{fig4} displays the constraint results on the parameters $\psi_{\rm GRB}(0)$, $a$, $b$, $L_{c,0}$, $z_c$, $n_1$, and $n_2$.
We find that the expectation from the empirical density case also provide a good representation of
the observed $z$ and $L$ distributions (green dotted lines in Figure~\ref{fig5}).
According to the AIC, the empirical density model is slightly favored compared
to the density evolution model, but the differences are statistically insignificant
($\sim 47\%$ for the former versus $\sim 43\%$ for the latter).
In both of these two models, the GRB formation rate is found to peak at a higher redshift
with respect to the no-evolution model. The intrinsic GRB formation rate as a function of redshift
in different models are displayed in Figure~\ref{fig6}. The comparison between the no-evolution
and empirical density models shows that the GRB rate may follow the SFR for $z\la2$,
but shows an enhancement compared to the SFR for $z\ga2$.

\begin{figure*}
  \centering
  % Requires \usepackage{graphicx}
  \includegraphics[scale=0.4]{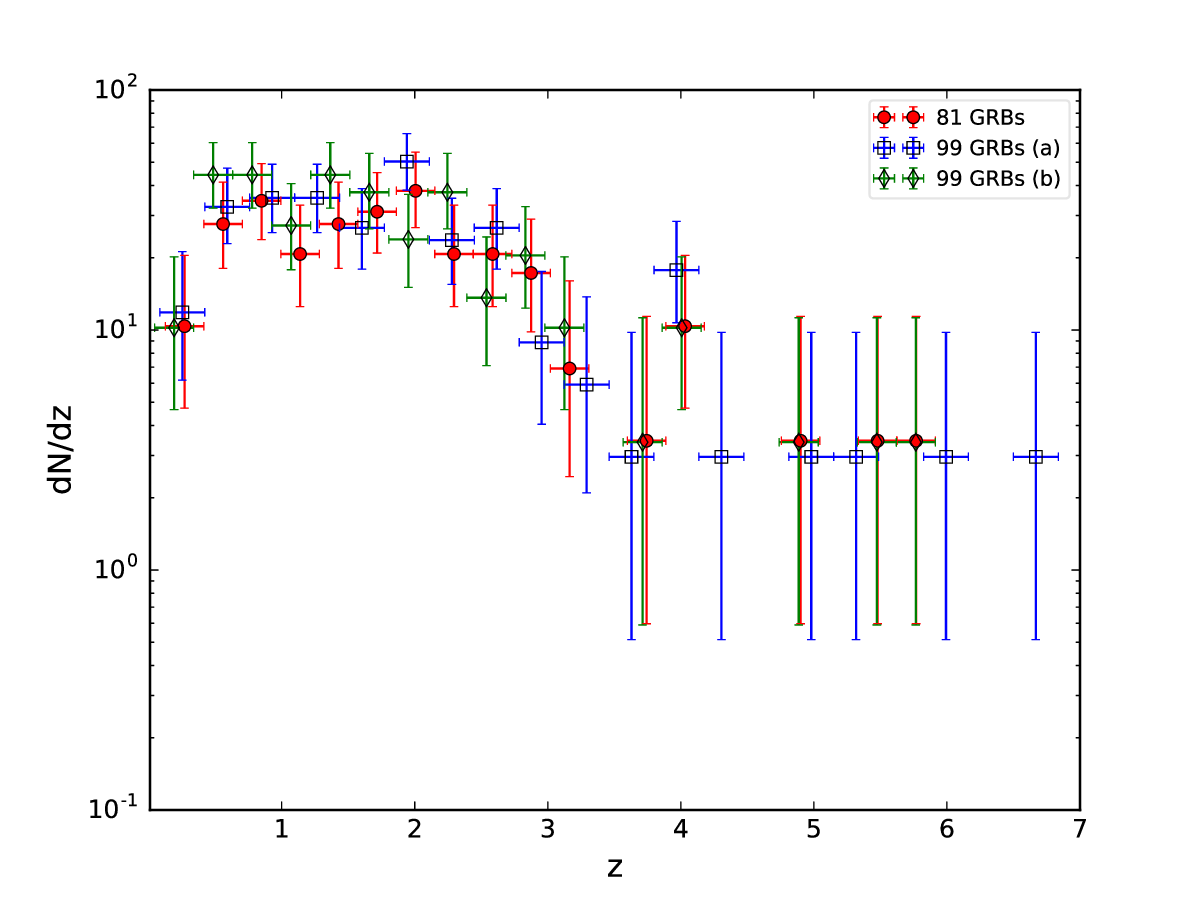}
  \includegraphics[scale=0.4]{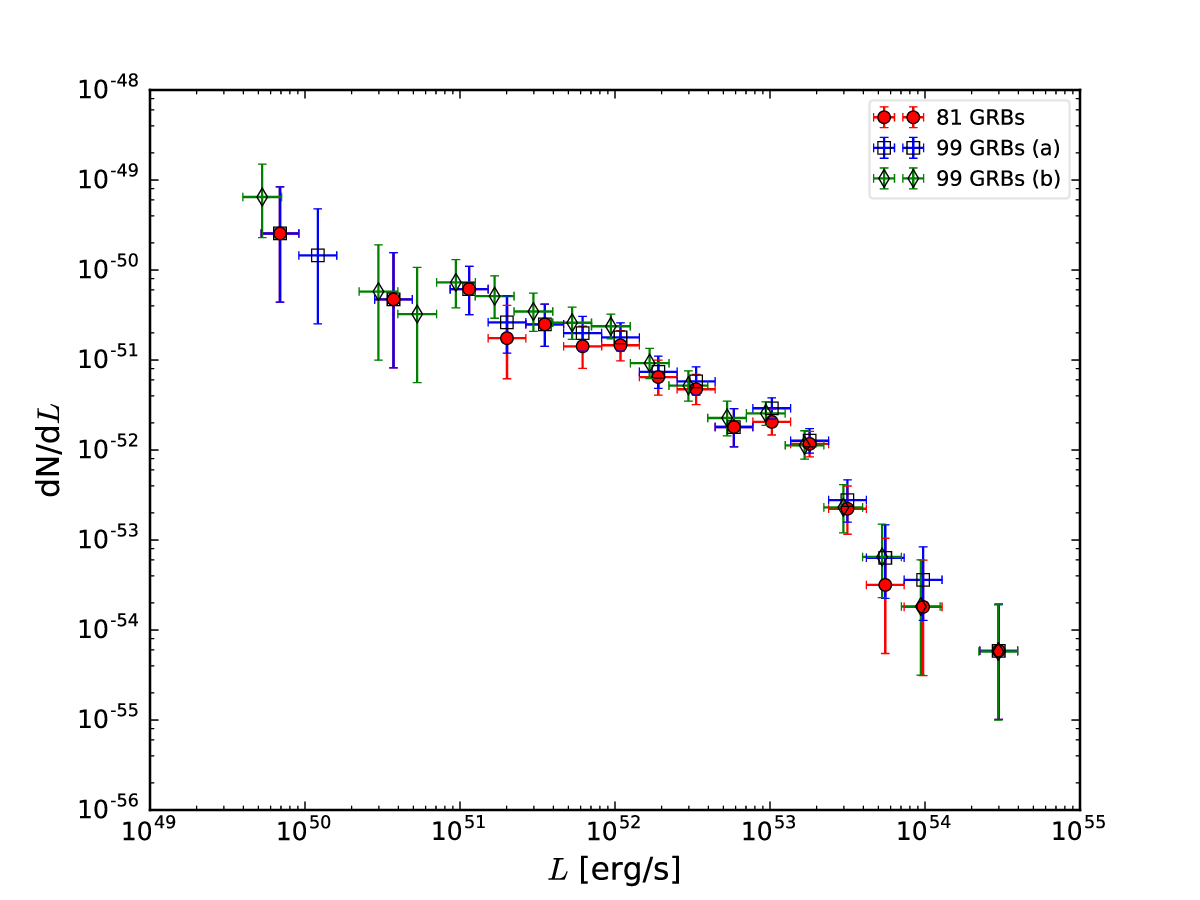}
  \caption{Redshift (left panel) and luminosity (right panel) distributions of 81 GRBs with measured redshifts (red dots)
  and of 99 GRBs (including 81 bursts with measured redshifts and 18 ones with pseudo redshifts), respectively.
  The pseudo redshifts of the 18 bursts without measured $z$ are estimated through the $E_{p}$--$L$ correlation.
  Blue squares and green diamonds correspond to the cases of requiring the 18 GRBs enter the $1\sigma$ and $2\sigma$ regions of
the correlation, respectively.}\label{fig7}
\end{figure*}

\section{Conclusion and Discussion}
\label{sec:summary}
In this work, we try to investigate the properties of long GRBs and their evolution with redshift.
To achieve this aim, a maximum likelihood method \citep{1983ApJ...269...35M} is adopted,
whose specific version has already been applied to GRBs \citep{2010MNRAS.406.1944W}.
Here we apply this method, for the first time, to a carefully selected sample of long GRBs
that is complete in peak flux and 82\% complete in redshift. This sample is composed of
GRBs detected by the $Swift$ satellite with favorable observing conditions for redshift
determination and with peak photon fluxes $P\ge 2.6 $ ph cm$^{-2}$ s$^{-1}$ \citep{2016A&A...587A..40P}.
It contains 99 bursts with a completeness of $\sim82\%$ (81 out of 99 bursts with known $z$ and $L$).

Based on the complete GRB sample, we directly construct the GRB luminosity function and redshift distribution
in the frameworks of different evolution models using the maximum likelihood method that
performs an analysis for the redshift and luminosity distributions of the GRB sample and
the MCMC technology that provides the best-fitting parameters (see Table~\ref{tab1})
and the probability density distributions of the parameters in each model (see Figures~\ref{fig1}-\ref{fig4}).
According to the AIC model selection criterion, we confirm that the no-evolution model can be safely excluded.
That is, GRBs must have experienced some kind of evolution to become more luminous or more population
in the past than present day. In order to account for the observed distributions, the GRB luminosity
should increase to $(1+z)^{2.22^{+0.32}_{-0.31}}$  or the GRB rate density should increase to $(1+z)^{1.92^{+0.20}_{-0.21}}$
with respect to the known cosmic SFR. But since the luminosity and density evolution scenarios predict very
similar distributions, we can not distinguish between these two kinds of evolutions simply on the basis of
the sample used in this study. These results are in good agreement with those of other works
\citep{2012ApJ...749...68S,2016A&A...587A..40P}.

Note that both the luminosity and density evolution models explored here are based on the assumption that
the GRB formation rate is related to a given SFR. With different SFR models, the constraint results
may change in some degree (see \citealt{2011MNRAS.417.3025V,2013ApJ...772...42H,2016JHEAp...9....1W}).
Therefore, we also consider an empirical density case in which the GRB rate density is
approximated as a broken power-law function, rather than being related to the SFR. In this case, we find that
the GRB rate rapidly increases at $z\la2$ ($n_1\simeq3.85$) and then shows slowly decreases at $z\ga2$ ($n_2\simeq-1.07$).
The GRB rate is compatible, of course, with a constant rate at $z\ga2$. The local formation rate of GRBs
is $1.49^{+0.63}_{-0.64}$ Gpc$^{-3}$ yr$^{-1}$, which is consistent with the result of \cite{2010MNRAS.406.1944W}.
By comparing the derived intrinsic redshift distributions in the no-evolution and empirical density models,
we find that while the GRB rate may be consistent with the SFR at $z\la2$, its high-redshift slope
is shallower than the steep decline of the SFR at $z\ga2$.

The total (complete) sample presented by \cite{2016A&A...587A..40P} comprises 99 GRBs.
However, only 81 of them with known redshift and luminosity have been used in our analysis.
To investigate the impact of the inclusion of the other 18 bursts without known redshift
on our conclusions, we apply the luminosity correlation \citep{2012MNRAS.421.1256N},
$\log \left[E_{p}(1+z)\right]=-25.33+0.53\log L$, to estimate their pseudo redshifts.
We use the observed peak flux and $E_{p}$ of the 18 bursts to calculate the rest-frame peak energies
and the isotropic peak luminosities for different redshifts. By requiring the bursts enter the
$2\sigma$ (or $1\sigma$) region of the correlation, we derive the lower limits of redshifts and then
the corresponding luminosities for these 18 bursts. In Figure~\ref{fig7}, we present the $z$ and
$L$ distributions of 81 GRBs with measured redshifts (red dots) and of 99 GRBs (including 81 bursts with
measured redshifts and 18 ones with pseudo redshifts), respectively. Blue squares and green diamonds
correspond to the cases of requiring the 18 GRBs enter the $1\sigma$ and $2\sigma$ regions of
the luminosity correlation, respectively. One can see from this plot that the $z$ and $L$ distributions
of these three cases are almost the same, we can therefore conclude that the inclusion of this 18\%
of bursts without known $z$ would not change the main conclusions of our paper.

\section*{Acknowledgements}
We are grateful to the anonymous referee for insightful comments.
We also thank Bin-Bin Zhang for helpful discussion on the average exposure factor of $Fermi$/GBM.
This work is partially supported by the National Natural Science Foundation of China
(grant Nos. 11603076, 11673068, 11725314, 11703094, and U1831122), the Youth Innovation Promotion
Association (2017366), the Key Research Program of Frontier Sciences (QYZDB-SSW-SYS005),
the Strategic Priority Research Program ``Multi-waveband gravitational wave Universe''
(grant No. XDB23000000) of Chinese Academy of Sciences, and the ``333 Project''
and the Natural Science Foundation (grant No. BK20161096) of Jiangsu Province.
%%%%%%%%%%%%%%%%%%%%%%%%%%%%%%%%%%%%%%%%%%%%%%%%%%

%%%%%%%%%%%%%%%%%%%% REFERENCES %%%%%%%%%%%%%%%%%%

% The best way to enter references is to use BibTeX:

%\bibliographystyle{mnras}
%\bibliography{ms} % if your bibtex file is called example.bib

% Alternatively you could enter them by hand, like this:
% This method is tedious and prone to error if you have lots of references
%\begin{thebibliography}{99}
%\bibitem[\protect\citeauthoryear{Author}{2012}]{Author2012}
%Author A.~N., 2013, Journal of Improbable Astronomy, 1, 1
%\bibitem[\protect\citeauthoryear{Others}{2013}]{Others2013}
%Others S., 2012, Journal of Interesting Stuff, 17, 198
%\end{thebibliography}

%%%%%%%%%%%%%%%%%%%%%%%%%%%%%%%%%%%%%%%%%%%%%%%%%%

% Don't change these lines
\bsp	% typesetting comment
\label{lastpage}
\end{document}